\documentclass{amsproc}
\usepackage{epsfig}
\newtheorem{theorem}{Theorem}[section]
\newtheorem{lemma}[theorem]{Lemma}

\theoremstyle{definition}

\theoremstyle{remark}

\numberwithin{equation}{section}

\begin{document}
\title{Quantum walks on graphs and quantum scattering theory}
\author{Edgar Feldman} 
\address{Department of Mathematics, Graduate Center of CUNY, 
365 Fifth Avenue, New York,NY 10016} 
\author{Mark Hillery}  
\address{Department of Physics, Hunter College ofCUNY, 695 Park Avenue, 
New York, NY 10021}
\thanks{The second author is supported by the National Science Foundation 
under grant 0139692.}
\subjclass{81U99,05c99}
\begin{abstract}
A number of classical algorithms are based on random walks on graphs.  It
is hoped that recently defined quantum walks can serve as the basis for
quantum algorithms that will faster than the corresponding classical ones. 
We discuss a particular kind of quantum walk on a general graph.  We affix
two semi-infinite lines to a general finite graph, which we call tails.
On the tails, the particle
making the walk simply advances one unit at each time step, so that its
behavior there is analogous to free propatation.  We are interested in how
many steps it will take the particle, starting on one tail and propagating
through the graph (where its propagation is not free), to emerge onto the
other tail.  The probability to make such a walk in $n$ steps and the
hitting time for such a walk can be expressed in terms of the transmission
amplitude for the graph, which is one element of its S matrix.  Demonstrating
this neccessitates a study of the analyticity properties of the transmission
and reflection coefficients of a graph.  We show
that a graph can have bound states that cannot be accessed by a particle
entering the graph from one of the tails.  Time-reversal invariance of a
quantum walk is defined and used to show that the transmission amplitudes
for the particle entering the graph from different directions are the same
if the walk is time-reversal invariant.  
\end{abstract}

\maketitle

\section{Introduction}
Random walks on graphs serve as the basis of algorithms for solving a
number of problems, including 2-SAT, graph connectivity, and finding 
satsifying assignments for Boolean functions. Quantum algorithms have
shown promise in solving some problems faster than is possible using
classical algorithms. Quantum walks represent an attempt to ``quantize''
classical random walk algorithms and thereby increase their speed.

Quantum algorithms must be run on a quantum computer.  In this kind of
machine, information is represented not in terms of bits, but in terms
of quantum bits, or qubits.  A bit is either $0$ or $1$ and can be
represented by current on or off, or by two different voltages.  A
qubit is a two-level quantum system, and possibilites include the spin
states of an electron or the polarization states of a photon.  One of the
states, e.g.\ spin up, corresponds to $0$ and the other, e.g.\ spin down,
corresponds to $1$.  The principle difference between a bit and a qubit, is
that the bit is either $0$ or $1$, while the qubit can be in a superposition
of its $0$ state and its $1$ state.  Computers based on bits represent
and process information according to the rules of classical physics, while
computers based on qubits represent and process information according to
the rules of quantum physics.

There has been some progress in using quantum walks to construct quantum
algorithms.  It has recently been shown that it is possible to
use a quantum walk to perform a search on the hypercube faster than
can be done classically \cite{shenvi}.  In this problem the number of
steps drops from $N$, which is the number of vertices, in the classical
case to $\sqrt{N}$ in the quantum case.  A much more dramatic 
improvement has recently been obtained by Childs, \emph{et al.} \cite{childs2}.
They constructed an oracle problem that can be solved by a quantum
algorithm based on a quantum walk exponentially faster than is possible
with any classical algorithm.

Quantum walks were introduced by Y.\ Aharanov, L.\ Davidovich and N.\ Zagury 
\cite{davidovich}.  They considered a spin-$1/2$ particle that is
shifted to the left or right depending on its spin state.  Since then,
a number of different kinds of quantum walks have been developed.  
The first, the coined walk, is a discrete time walk that makes use of 
an auxiliary quantum system, the coin.  This is necessary to make the 
operator that advances the walk one step unitary.  This walk was 
originally proposed by Watrous and is closlely related to the original walk
discussed by Aharanov \cite{watrous}.  A second type of walk is the continuous
time walk, due to Childs, \emph{et al}. \cite{childs}.  Here the
Hamiltonian that is used to construct the time development transformation
is taken to be proportional to the adjancency matrix of the graph.  A
third type of walk, based on thinking about the graph as an interferometer
with optical multiports as the nodes was recenty proposed by us 
\cite{hillery}.  In this case the walk takes place on the edges of the
graph, rather than the vertices, and each edge has two states, one
corresponding to traversing the edge in one direction and the other to
traversing the edge in the opposite direction.  It should be noted that
while it is very simple to construct a quantum walk on an arbitrary
graph using the second and third methods, it is much less obvious how to 
do so for the coined walk.

Let us briefly describe the coined quantum walk. In trying to formulate
a quantum walk on a graph, the most natural thing to do is to let a
set of orthonormal basis states correspond to the vertices of the graph.
If a particle is in the state $|n\rangle$, that corresponds to its being
located on vertex $n$.  Let the Hilbert space spanned by these states
be $\mathcal{H}_{p}$. Trying to define a unitary evolution using this
scheme soon leads to serious problems, as was first noted by Meyer
\cite{meyer}.  Watrous solved this problem by enlarging the Hilbert
space in which the quantum walk takes place.  How this scheme works
is most easily seen by considering the quantum walk on a line.  The
vertices are labelled by integers, and, in addition, there is a quantum
coin, which has two states, $|L\rangle$ and $|R\rangle$, corresponding
to left and right, respectively, which span a two-dimensional Hilbert
space, $\mathcal{H}_{c}$.  A basis for the Hilbert space describing
the entire system, $\mathcal{H}_{p}\otimes\mathcal{H}_{c}$, 
 is given by the states $|n\rangle\otimes |\alpha\rangle$, where 
$n$ is an integer, and $\alpha$ is either $L$ or $R$.  A step in this walk 
consists of applying the Hadamard operator, $H$, to the coin,
\begin{eqnarray}
H|L\rangle & = & \frac{1}{\sqrt{2}}(|L\rangle + |R\rangle ) \nonumber \\  
H|R\rangle & = & \frac{1}{\sqrt{2}}(|L\rangle - |R\rangle ) .
\end{eqnarray}
and then the operator 
\begin{equation}
V_{H} = S\otimes |R\rangle\langle R|+ S^{\dagger}\otimes |L\rangle\langle L|,
\end{equation}
where $S$ is the shift operator, whose action is given by
\begin{equation}
S|n\rangle = |n+1\rangle \hspace{1cm} S^{\dagger}|n\rangle = |n-1\rangle .
\end{equation}
Note that the evolution of the quantum state of the walk is deterministic,
that is, if the initial state is $|\psi_{in}\rangle$, then the state after
$m$ steps is just $[V_{H}(I_{p}\otimes H)]^{m}|\psi_{in}\rangle$.
Probabilities enter the picture when we try to determine where the particle
is by making measurements.

Coined quantum walks have been studied on both the line by Nayak 
and Vishwananth \cite{nayak} 
and on the cycle by D.\ Aharanov \emph{et al}. \cite{aharonov}.
Aharonov \emph{et al}. also studied a number of properties of quantum
walks on general graphs.  Absorbing times and probabilities for walks on
the line have been examined by several authors \cite{yamasaki,bach}.
Going beyond one dimension, properties of quantum walks on two and three 
dimensional lattices \cite{mackay} and on the hypercube \cite{moore,kempe}
have been explored. The effects of decoherence on quantum walks have also 
been studied.  Brun, \emph{et al}. showed how increasing amounts of
decoherence turn a quantum walk into a classical random walk \cite{brun}.
Kendon and Tregenna found that small amounts of decoherence can speed up
the convergence of the time-averaged probability distribution of a 
quantum walk \cite{kendon}.  Many aspects of quantum walks are covered in
the recent and excellent review by Kempe \cite{kemperev}. 

Several proposals have been made for the physical realization of quantum
walks \cite{travaglione}-\cite{knight2}.  The ones closest in spirit to the
walks studied here are the realizations that employ optical methods,
either linear optical elements \cite{zhao,jeong} or cavities 
\cite{knight1,knight2}.  The last two references show that an experimental
quantum walk has, in fact, been carried out, though it was not intrepreted
as such at the time \cite{bouwmeester}.

Here we wish to describe quantm walks that take place on the edges of a
graph, and to study the connection between these walks and scattering
theory.  In order to show how scattering theory enters the picture, begin 
with a graph, choose two vertices, and attach two 
semi-infinite lines, which we shall call tails, to it, one to each of the 
selected vertices.  Each of the half-lines is made up of vertices and edges.  
Thus we can start the walk on one of the tails, have it progress into the
original graph, and emerge onto the opposite tail.  This type of 
arrangement allows us to define a scattering matrix, or S matrix, 
for the original graph, with
the amplitude to get from one tail to the other being called the 
transmission amplitude.  As
we shall see properties of quantum walks starting on one tail and ending
on the other can be expressed in terms of the transmission amplitude of the
graph.  It should be noted that scattering approaches have also proven 
useful in a related area, quantum graphs, which can be used to study 
electron transport in large molecules \cite{schmidt}.

The first use of scattering theory to analyze the behavior of quantum walks 
was done by Farhi and Gutmann \cite{farhi}.  They studied continuous-time
walks on trees, and they added tails to the trees to place the walks into a
scattering-theoretic framework.  They were able to place bounds on the time
for a walk, starting at the root of the tree, to penetrate to one of the 
leaves.

We begin by defining our quantum walk, and then illustrate it with an
example.  We then explore the relation between properties of the quantum
walk and the transmission amplitude of the graph.  These are then
applied to an example.  We define a time-reversal transformation for
quantum walks and use it to explore when the transmission amplitude to
go through the graph in one direction is the same as the transmission
amplitude to go through it in the other direction.  Finally, we
finish with a detailed study of the properties of the transmission and
reflection amplitudes, in particular showing that they are analytic in
on the unit circle and in its interior.  This allows us to express the
probability of starting on one tail and finishing on the other in $n$
steps and also the hitting time for a walk from one tail to the other
very simply in terms of the transmission amplitude of the graph. 

\section{Quantum walks on edges}
The type of quantum walk that will be used throughout
this paper was originally presented in \cite{hillery}.  We imagine
a particle on an edge of a graph; it is this particle that will make the
walk.  Each edge has two states, one going in one direction, the other
going in the other direction, and we denote the set of oriented edges
(an edge and its direction of traversal is an oriented edge) of the graph 
by $E$.  That is, if our edge is between the
vertices $A$ and $B$, which we shall denote as $(A,B)$, it has two
orthogonal states, $|A,B\rangle$, corresponding to the particle being
on $(A,B)$ and going from $A$ to $B$, and $|B,A\rangle$, corresponding to 
the particle being on $(A,B)$ and going from $B$ to $A$.  The collection
of all of these edge states is a basis for a Hilbert space, and the states
of the particle making the walk lie in this space, which is $L^{2}(E)$.

Now that we have our state space, we need a unitary operator that advances
the walk one step.  Let us first illustrate how this works for a walk
on the line.  We shall label the vertices by the integers.  In this case,
the states of the system are $|j,k\rangle$, where $k=j\pm 1$. 
The vertices can be thought of as scattering centers.  Consider what happens
when a particle, moving in one dimension, hits a scattering
center.  It has a certain amplitude to continue in
the direction it was going, i.e.\ to be transmitted, and an amplitude
to be change its direction, i.e.\ to be reflected.  The scatterer
has two input states, the particle can enter from either the
right or the left, and two output states, the particle can leave heading
either right or left.  The scattering center defines a unitary transformation
between the input and output states.

We now need to translate this into transition rules for our
quantum walk.  Suppose we are in the state $|j-1,j\rangle$.  If the
particle is transmitted it will be in the state $|j,j+1\rangle$, and if
reflected in the state $|j,j-1\rangle$.  Let the transmission amplitude
be $t$, and the reflection amplitude be $r$.  We then have the
transition rule
\begin{equation}
|j-1,j\rangle\rightarrow t|j,j+1\rangle +r|j,j-1\rangle ,
\end{equation}
where unitarity implies that $|t|^{2}+|r|^{2}=1$.
The other possibility is that the particle is incident on vertex
$j$ from the right, that is it is in the state $|j+1,j\rangle$.
If it is transmitted it is in the state $|j,j-1\rangle$, and if
it is reflected, it is in the state $|j,j+1\rangle$.  Unitarity
of the scattering transformation then gives us that
\begin{equation}
|j+1,j\rangle\rightarrow t^{\ast}|j,j-1\rangle -r^{\ast}|j,j+1\rangle .
\end{equation}
These rules specify our walk.

The case $t=1$ and $r=0$ corresponds to free particle propagation; a
particle in the state $|j,j+1\rangle$ simply moves one step to the
right with each time step in the walk.  If $r\neq 0$, then there is
some amplitude to move both to the right and to the left.  As an example,
let us consider the case when $t=r=1/\sqrt{2}$, and the particle starts
in the state $|0,1\rangle$.  The probability distribution for finding the
particle on a given edge after $1,000$ steps is shown in Figure 1. First,
note that this distribution look nothing like the distribution that would
result from a classical random walk on a line, which would be a Gaussian
centered about the initial position of the particle.  In addition, we
see that there is an interval about the origin in which it is very likely
to find the particle.  An asymptotic analysis of the walk shows that
this region lies between $-|t|n$ and $|t|n$, where $n$ is the number of
steps in the walk and $t$ is the transmission amplitude of the vertices
\cite{hillery}.  That is, the region in which the the particle is likely 
to be found grows as $n$ instead of $\sqrt{n}$ as would be the case in a 
standard random walk. 

\begin{figure}
\label{fig1}
\epsfig{file=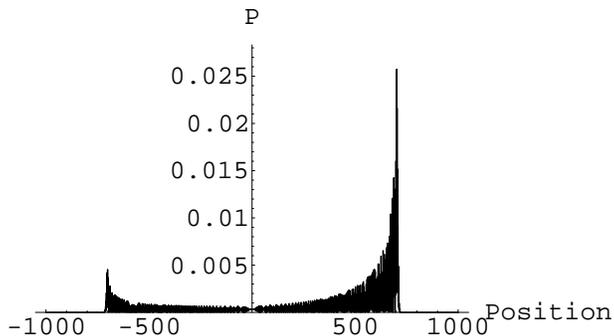}
\caption{Probability distribution for quantum walk after
1000 steps}
\end{figure}

Thinking about the particle making the walk and the scattering centers
as beam splitters, we see that this type of quantum walk is analogous
to the propagation of light through an interferometer.  This suggests
that we can add a
new element to quantum walks that has no analogue in classical random
walks.  Interferometers are made up of multiports (generalized beam
splitters) \cite{zeilinger} and phase shifters;
a phase shifter imparts a constant phase to a photon that passes through
it.  Suppose we were to put a phase shifter that imparts a phase shift of
$\phi$ just before the $j^{\rm th}$ vertex.  The transition rules for
the states adjacent to this vertex are modified, while the rules for
all other states are unaffected.  In particular, we now have
\begin{eqnarray}
|j-1,j\rangle &\rightarrow & te^{i\phi}|j,j+1\rangle +re^{2i\phi}
|j,j-1\rangle \nonumber \\
|j+1,j\rangle &\rightarrow & -r^{\ast}|j,j+1\rangle +t^{\ast}
e^{i\phi}|j,j-1\rangle .
\end{eqnarray}
Insertion of a phase shifter into an edge can change the properties
of a quantum walk, because it changes how different paths interfere.  

So far we have only considered vertices at which two edges meet, but
if we are to construct graphs more complicated than lines, we need
to see how a vertex with more that two edges emanating from it behaves.
We shall begin by looking at an example.

If a vertex treats all edges entering it in an equivalent fashion, then we
have a particulary simple situation, because the edges of the graph do not
have to be labelled.  One vertex of this type is very closely related to 
the quantum coin used in the
walk on the hypercube \cite{moore,kempe}.  Let the vertex at which
all of the edges meet be labelled by $O$, and the opposite ends of
the edges be labelled by the numbers $1$ through $n$.  For any 
input state, $|k,O\rangle$, where $k$ is an integer between $1$ and
$n$, the transition rule is that the amplitude to go the output
state $|O,k\rangle$ is $r$, and the amplitude to go to any other
output state is $t$.  That is, the amplitude to be reflected is $r$,
and the amplitude to be transmitted through any of the other edges
is $t$.  Unitarity places two conditions on these amplitudes
\begin{eqnarray}
(n-1)|t|^{2}+|r|^{2}=1 \nonumber \\
(n-2)|t|^{2}+r^{\ast}t+t^{\ast}r =0 .
\end{eqnarray}
We shall call such vertices equal-transmission vertices.
As an example, for the case $n=3$, possible values of $r$ and $t$
are $r=-1/3$ and $t=2/3$.  
 
In order to construct a walk for a general graph, one chooses a unitary
operator for each vertex, i.e.\ one that maps the states coming into
a vertex to states leaving the same vertex.  More precisely, let $\tau_{A}$
be the set of oriented edges starting at the vertex $A$ and $T_{A}$ be
the linear span of $\tau_{A}$.  Similarly, let $\omega_{A}$ be the set
of oriented edges ending at the vertex $A$, and $\Omega_{A}$ be the
linear span of $\omega_{A}$.  We have that $\tau_{A}\bigcap \tau_{B}=
\emptyset$ and $\omega_{A}\bigcap \omega_{B} =\emptyset$ if $A\neq B$,
which implies that 
\begin{equation}
E=\bigcup_{A}\tau_{A}=\bigcup_{A}\omega_{A} ,
\end{equation}
and
\begin{equation}
L^{2}(E)=\bigoplus_{A} T_{A}= \bigoplus_{A} \Omega_{A},
\end{equation}
where the unions and direct sums are over all vertices of the graph.
The local unitary operator, $U_{A}$ maps $\Omega_{A}$ to $T_{A}$, and 
describes the scattering at vertex $A$.  One step of the walk consists
of the combined effect of all of these operations; the overall unitary
operator, $U$, that advances the walk one step is constructed from the local
operators for each vertex.  Explicitly, the edge state $|A,B\rangle$,
which is the state for the particle going from vertex $A$ to vertex $B$, will
go to the state $U_{B}|A,B\rangle$ after one step, where $U_{B}$ is
the operator corresponding to vertex $B$.  This prescription guarantees
that the overall operation is unitary, in particular, $U$ acting on
any other edge state $|C,D\rangle$ will give a state orthogonal to
$U|A,B\rangle$.  If $D=B$, then $|A,B\rangle$ and $|C,D\rangle$ will be
mapped into $T_{B}$,
but the unitarity of $U_{B}$ ensures that $U|A,B\rangle$ and
$U|C,D\rangle$ are orthogonal.  If $D\neq B$, then $U$ maps $|A,B\rangle$
and $|C,D\rangle$ onto different sets of states, $T_{B}$ and $T_{D}$,
respectively, and the results are
then orthogonal.  Therefore, as the edge states make up an orthonormal
basis of the Hilbert space in which the walk occurs, and $U$ maps 
this basis to another orthonormal basis, it is unitary.

\section{Example}
Let us put all of this together in a very simple example.  Consider the
graph shown in Figure 2, 
where each of the vertices where two edges meet have $t=1$ and $r=0$,
while the three-edge vertices are of the type discussed in the
previous section, with $r=-1/3$ and $t=2/3$.  There is a phase
shifter with an adjustable phase, $\phi$, in one of the edges.  The graph
goes to negative infinity on the left and plus infinity on the right.
Note that it is very simple to construct the unitary operator that
advances a quantum walk on this graph by one step; one simply combines
the actions of the operators corresponding to each vertex. 

\begin{figure}
\label{fig2}
\epsfig{file=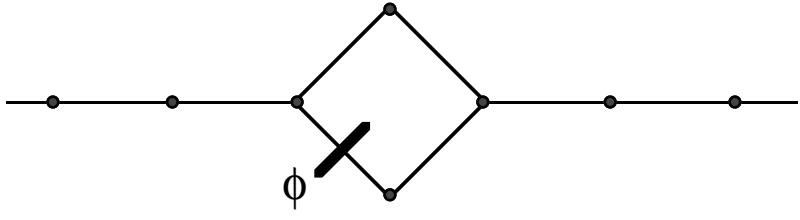}
\caption{Graph for example}
\end{figure}

We can find the unnormalized eigenstates for this graph, and one set of
them can be described as having an incoming wave from the left, an
outgoing transmitted wave going to the right, and a reflected wave
going to the left. A second set will have an incoming wave from the right,
an outgoing trasmitted wave to the left, and a refelected wave to the
right.  The mathematical status of these objects will be explored in
Sections 7 and 8.
Finally, there may be bound states, i.e.\ eigenstates that are
localized in the region between the two vertices with three edges.
Now, let us denote the left three-edge
vertex by $0$ and the right one by $2$, and number the vertices on the
lines correspondingly, from $2$ to plus infinity to the right and
from $0$ to minus infinity to the left.  The eigenstates with a wave
incident from the left take the form
\begin{eqnarray}
|\Psi\rangle & = &  \sum_{j=-\infty}^{-1}(e^{ij\theta}|j,j+1\rangle +
b_{-1}e^{-i(j+1)\theta}|j+1,j\rangle )+|\Psi_{02}\rangle \nonumber \\
 & &+\sum_{j=2}^{\infty}a_{2}e^{i(j-2)\theta}|j,j+1\rangle ,
\end{eqnarray} 
where $|\Psi_{02}\rangle$ is the part of the eigenfunction between 
vertices $0$ and $2$, and $e^{-i\theta}$ is the eigenvalue of the
operator $U$ that advances the walk one step.  The first term can be thought
of as the incoming wave; it is confined to the region between negative
infinity and $0$, and consists of states in which the particle is moving
to the right.  The term proportional to $b_{-1}$ is the reflected wave.
It is also confined to the region between negative infinity and $0$, but
consists of states in which the particle is moving to the left.  Finally,
the term proportional to $a_{2}$ is the transmitted, being confined to
the region from $2$ to infinity, and consisting of states with the particle
moving to the right. The coefficient $b_{-1}$ can be interpretted as the
amplitude for the incoming wave to be reflected, and $a_{2}$ as the
amplitude to be transmitted.  Denoting the upper
vertex between $0$ and $2$ as $1A$ and the lower one as $1B$, we
can express $|\Psi_{02}\rangle$ as
\begin{eqnarray}
|\Psi_{02}\rangle & = & a_{0A}|0,1A\rangle + a_{1A}|1A,2\rangle + 
b_{0A}|1A,0\rangle + b_{1A}|2,1A\rangle \nonumber \\
 & & +a_{0B}|0,1B\rangle + a_{1B}|1B,2\rangle + b_{0B}|1B,0\rangle
+b_{1B}|2,1B\rangle .
\end{eqnarray}
The solution of the equation $U|\Psi\rangle = e^{-i\theta}|\Psi\rangle$
is given in the appendix.

One can define a transmission coefficient for this 
graph, which is just the ratio of the intensities of the outgoing
transmitted and the incoming waves,
\begin{equation}
T=|a_{2}|^{2}=\left|\frac{4(1+e^{-i\phi})(1-e^{-i(4\theta +\phi)})}
{e^{-4i\theta}(1+e^{-i\phi})^{2}-(3e^{-i(4\theta +\phi)}-1)^{2}}
\right|^{2} .
\end{equation}
One finds that for $\phi =0$, T is nonzero, because the two paths 
from $0$ to $2$ interfere constructively, while when $\phi =\pi$
they interfere destructively, which results in $T=0$.  Therefore,
the behavior of the walk strongly depends on the value of the
phase shift.  

\section{Hitting times}
Suppose we want to find out whether the particle making the walk is on
the edge between vertices $j$ and $j+1$.  The relevant projection
operator is
\begin{equation}
P_{j}=|j,j+1\rangle\langle j,j+1| + |j+1,j\rangle\langle j+1,j| .
\end{equation}
If we obtain $1$ the particle is on that edge, and if we obtain $0$, it
is not.  What we wish to find out is, if the particle is initially in
the state $|\Psi (0)\rangle$, the probability that the particle is not
on the edge for the first $n$ steps of the walk, but is on the $(n+1)$st.
We measure $P_{j}$ after every step, and we have to take the effect of
these meaurements into account in describing the dynamics of the wave 
function.

Let us first see what happens after one step.  Let $U$ be the unitary 
operator that advances the walk one step.  The probability that the 
particle is not on the edge between $j$ and $j+1$ after one step, $p(1)$, is
\begin{equation}
p(1)=\| (I-P_{j})U\Psi (0)\|^{2} .
\end{equation}
and the quantum state, assuming the particle was not found on this edge,
is
\begin{equation}
|\Psi (1)\rangle = \frac{(I-P_{j})U|\Psi (0)\rangle }{\| (I-P_{j})
U\Psi (0)\| }  .
\end{equation}
Now let us go one more step to see the pattern.  The probability that
the particle is not on the edge between $j$ and $j+1$ after either
the first or second steps, $p(2)$, is
\begin{eqnarray}
p(2) & = & p(1) \| (I-P_{j})U\Psi (1)\|^{2} \nonumber \\
 & = & \| (I-P_{j})U(I-P_{j})U\Psi (0)\|^{2} ,
\end{eqnarray}
and the quantum state, assuming the particle was not found on the edge
after either step, is
\begin{equation}
|\Psi (2)\rangle = \frac{(I-P_{j})U|\Psi (1)\rangle }{\| (I-P_{j})
U\Psi (1)\| }  .
\end{equation}
In general, if $p(n)$ is the probability of not finding the particle
on the edge between $j$ and $j+1$ after each of $n$ steps, then
\begin{equation}
\label{probn}
p(n)= p(n-1) \| (I-P_{j})U\Psi (n-1)\|^{2} ,
\end{equation}
where $|\Psi (n-1)\rangle$ is the quantum state after $n-1$ steps 
and $n-1$ measurements indicating that the particle was not on the
designated edge.  We also have that
\begin{equation}
\label{staten}
|\Psi (n)\rangle = \frac{(I-P_{j})U|\Psi (n-1)\rangle }{\| (I-P_{j})
U\Psi (n-1)\| }  .
\end{equation}

Let us now proceed by induction.  We shall assume that
\begin{eqnarray}
p(n)= \| [(I-P_{j})U]^{n}\Psi (0)\|^{2} \nonumber \\
|\Psi (n)\rangle = \frac{[(I-P_{j})U]^{n}|\Psi (0)\rangle }
{\| [(I-P_{j})U]^{n}\Psi (0)\| } ,
\end{eqnarray}
which clearly holds for $n=1,2$.  Substitution of the second of these
equations into Eq.\ (\ref{staten}) yields 
\begin{equation}
|\Psi (n+1)\rangle = \frac{(I-P_{j})U|\Psi (n)\rangle }{\| (I-P_{j})
U\Psi (n)\| }  ,
\end{equation}
so that part of our induction hypothesis is verfied.  From Eq.\ 
(\ref{probn}) we find that
\begin{eqnarray}
p(n+1) & = & \| [(I-P_{j})U]^{n}\Psi (0)\|^{2} \| (I-P_{j})U
\Psi (n)\|^{2} \nonumber \\
 & = & \| [(I-P_{j})U]^{n+1}\Psi (0)\|^{2} ,
\end{eqnarray}
which proves the remaining part.  This result is easily generalized to the
case of making measurements to determine whether the particle is on a
set of edges instead of on a single edge.  A similar result for the coined
quantum walk was derived by Bach, \emph{et al}. \cite{bach}.

The hitting time for a random walk on a graph is the expected number
of steps in a walk that starts at one specified vertex and stops on
first reaching a second specified vertex.  In order to calculate this
we need the probability that a walk starting on the first vertex does
not reach the second for its first $n-1$ steps, but does reach it
on the $n$th step.  The corresponding quantity for a quantum walk,
which we shall call $q(n)$, is the probability that a walk starting
in the state $|\Psi (0)\rangle$ is not measured to be on the edge
between $j$ and $j+1$ for the first $n-1$ steps, but is measured to
be on that edge at the $n$th step.  This probability is given by
\begin{eqnarray}
\label{qn}
q(n) & = & p(n-1)\|P_{j}U\Psi (n-1)\|^{2} \nonumber \\
 & = & \| P_{j}U[(I-P_{j})U]^{n-1}\Psi (0)\|^{2} .
\end{eqnarray}
The particle making this walk may never be found to be on the edge between 
$j$ and $j+1$ at all, and the probability that it is, is given by  
\begin{equation}
P_{out}=\sum_{n=1}^{\infty}q_{n} .
\end{equation}
We can define a conditional hitting time for this walk, $h$, as
\begin{equation}
\label{hit}
h=\frac{1}{P_{out}}\sum_{n=1}^{\infty}nq_{n}.
\end{equation}
This quantity provides an an indication of how many steps a walk that does
reach the edge between $j$ and $j+1$ will require.  
 
Let us now consider the situation depicted in Figure 3.
We start the walk on the edge $|-1,0\rangle$ moving to the right,
and we are interested in how long it takes to get to the edge
$|j,j+1\rangle$, where $G$ is a general graph with a finite number of 
edges and vertices.  The vertices outside of the graph, $G$, for example, 
$-1$ and $j+1$, have a transmission amplitude, $t=1$.  We shall denote
the entire graph, $G$ plus the tails, as $\tilde{G}$.  
The case considered in the previous section is
a special case of this situation.  When calculating the probability,
$q(n)$, that the walk arrives at the edge $|j,j+1\rangle$ for the
first time after $n$ steps, we need to evaluate (see Eq.\ (\ref{qn}))
\begin{equation}
\label{out}
P_{j}U[(I-P_{j})U]^{n-1}=P_{j}U^{n}-\sum_{k=1}^{n-1}P_{j}U^{(n-k)}
P_{j}U^{k} +\ldots
\end{equation}
We first note that once the particle leaves the graph, $G$, it moves
one unit to the right with each step.  That means that terms
of the form $P_{j}U^{m}P_{j}U^{k}|-1,0\rangle$, for $m>0$,  are 
zero, because the first 
$P_{j}$ puts the particle in the state $|j,j+1\rangle$, and the 
operator $U^{m}$ moves it $m$ steps to the right, so that the
particle in now in the state $|j+m,j+m+1\rangle$.  The projection
$P_{j}$ acting on this state gives zero.  Therefore, only the
first term on the right-hand side of Eq.\ (\ref{out}) gives a
nonzero contribution.  That means that in this case, we have
\begin{equation}
q(n)=\langle\Psi (0)|(U^{\dagger})^{n}P_{j}U^{n}|\Psi (0)\rangle .
\end{equation}

\begin{figure}
\label{fig3}
\epsfig{file=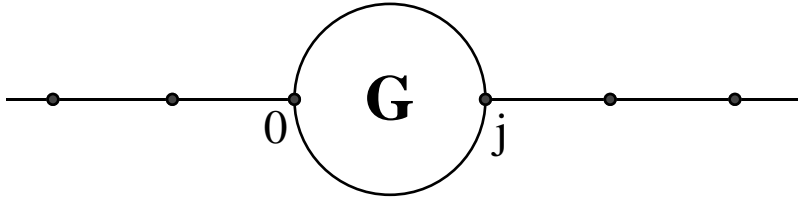}
\caption{A general graph, $G$, with tails attached to two if its vertices}
\end{figure}

We now want to relate the probability $q(n)$ to the transmission
coefficient of the graph, which we shall define shortly.  The eigenstates
of the system can be grouped into three sets, those with the particle
coming in from the left, those with the particle coming in from the right,
and bound states.
The eigenstates of this system corresponding
to the particle coming in from the left are
\begin{eqnarray}
\label{geneigen}
|\Psi_{l} (\theta )\rangle & = & \sum_{k=-\infty}^{-1}(e^{i(k+1)\theta}
|k,k+1\rangle +r(\theta )e^{-i(k+1)\theta}|k+1,k\rangle )+|\Psi_{G}
(\theta )\rangle \nonumber \\ 
 & & +\sum_{k=j}^{\infty}t(\theta )e^{i(k-j)\theta }
|k,k+1\rangle ,
\end{eqnarray}
where $r(\theta )$ is the reflection coefficient, $t(\theta )$ is the
transmission coefficient, and $|\Psi_{G}(\theta )\rangle$ is the 
part of the eigenfunction inside the graph $G$.  This state is, in fact,
not in $L^{2}(E)$, and its mathematical significance will be clarified
in a later section.  In Appendix B we show that 
\begin{equation}
\label{reftrans}
|r(\theta )|^{2}+|t(\theta )|^{2} = 1.
\end{equation}

For the initial state of our walk we will choose $|-1,0\rangle$, and we
want to express this state in terms of the eigenstates in Eq.\ 
(\ref{geneigen}).  To do so we integrate $|\Psi_{l} (\theta )\rangle$
from $0$ to $2\pi$, finding
\begin{eqnarray}
\frac{1}{2\pi}\int_{0}^{2\pi}d\theta |\Psi_{l} (\theta )\rangle & = & 
|-1,0\rangle +\frac{1}{2\pi}
\int_{0}^{2\pi}d\theta [\sum_{k=-\infty}^{-1}r(\theta )e^{-i(k+1)\theta}
|k+1,k\rangle ] \nonumber \\ 
 & & +\frac{1}{2\pi}\int_{0}^{2\pi}d\theta [|\Psi_{G}(\theta )\rangle +
\sum_{k=j}^{\infty}t(\theta )e^{i(k-j)\theta } |k,k+1\rangle ].
\end{eqnarray}
By setting $z=\exp (i\theta)$ the integrals on the right-hand side of the 
above equation can be turned into integrals along the unit circle, which we
shall denote by $C$, in the complex plane.  These integrals are of the form
\begin{eqnarray}
\label{int0}
\int_{C}dz\, z^{m-1}r(z) & \int_{C}dz\, \frac{1}{z}|\Psi_{G}(z)\rangle 
\nonumber \\
\int_{C}dz\, z^{m-1}t(z) & ,
\end{eqnarray}
where $m\geq 0$.  As we shall show in Section 7, $r(z)$, $t(z)$, and
$|\Psi_{G}(z)\rangle$ are analytic functions of $z$ on the unit circle
and in its interior, and they all vanish when $z=0$.  This implies that
all of the above integrals are zero.  Therefore, we can conclude that
\begin{equation}
|-1,0\rangle = \frac{1}{2\pi}\int_{0}^{2\pi}d\theta |\Psi_{l} (\theta )
\rangle .
\end{equation}

Now we have that
\begin{equation}
q(n) = |\langle j,j+1|U^{n}|-1,0\rangle |^{2} ,
\end{equation}
and
\begin{eqnarray}
\langle j,j+1|U^{n}|-1,0\rangle & = & \frac{1}{2\pi}\int_{0}^{2\pi}d\theta
e^{-in\theta}\langle j,j+1|\Psi_{l} (\theta )\rangle \nonumber \\
 & = &  \frac{1}{2\pi}\int_{0}^{2\pi}d\theta e^{-in\theta} t(\theta ) .
\end{eqnarray}
Therefore, we find
\begin{equation}
\label{qn2}
q(n)=\left| \frac{1}{2\pi}\int_{0}^{2\pi}d\theta e^{-in\theta} t(\theta ) 
\right|^{2} ,
\end{equation}
which relates a walk probability, the probability to land on the edge
between $j$ and $j+1$ for the first time after $n$ steps, to the S
matrix of the graph $G$.

We can rewrite the integral appearing in the above expression for $q(n)$ 
as a contour integral 
over the unit circle, $C$ by setting $z=e^{i\theta}$.  This gives us
\begin{equation}
\label{integral}
\frac{1}{2\pi}\int_{0}^{2\pi}d\theta e^{-in\theta} t(\theta ) =
\frac{1}{2\pi i}\int_{C}dz \frac{1}{z^{n+1}} t(z ).
\end{equation}
Because, as we shall see, $t(z)$ is analytic in a region containing the 
unit circle and its interior, 
the only contribution to the integral in Eq.\ (\ref{integral}) comes
from the pole at the origin that is due to the $1/z^{n+1}$ factor.  
This fact makes it straightforward to evaluate the above 
integral, by just finding the residue of the pole at the origin, giving us
\begin{equation}
\frac{1}{2\pi i}\int_{C}dz \frac{1}{z^{n+1}} t(z )
=\left. \frac{1}{n!}\frac{d^{n}}{dz^{n}}t(z)\right|_{z=0} .
\end{equation}
This is just the
$n^{\rm th}$ coefficient in the Taylor expansion of $t(z)$ about the 
origin, i.e.\ if
\begin{equation}
t(z)=\sum_{n=0}^{\infty}c_{n}z^{n} ,
\end{equation}
then we have that $q(n)=|c_{n}|^{2}$.  The conditional hitting time, 
defined in Eq.\ (\ref{hit}), is then given by
\begin{equation}
h=\frac{1}{P_{out}}\sum_{n=0}^{\infty}n|c_{n}|^{2} \hspace{1cm}
P_{out}=\sum_{n=0}^{\infty}|c_{n}|^{2} .
\end{equation}
We now note that 
\begin{equation}
t^{\ast}(z)z\frac{dt}{dz}=\sum_{m=0}^{\infty}\sum_{n=1}^{\infty}n
c^{\ast}_{m}c_{n}(z^{\ast})^{m}z^{n} ,
\end{equation}
from which it follows, setting $z=\exp (i\theta )$ and integrating,
that
\begin{equation}
h=\frac{1}{2\pi P_{out}}\int_{0}^{2\pi}d\theta t^{\ast}(e^{i\theta})e^{i\theta}
\left. \frac{dt}{dz}\right|_{z=e^{i\theta}} .
\end{equation}
Similarly, we have that
\begin{equation}
P_{out}=\frac{1}{2\pi}\int_{0}^{2\pi}d\theta\, |t(e^{i\theta })|^{2} .
\end{equation}
The integrals in the previous two equations can also be expressed as
contour integrals if we define the ``reflected'' transmission amplitude,
$t_{R}(z)$ to be $t_{R}(z)=t(z^{\ast})^{\ast}$.  We then have that
\begin{eqnarray}
P_{out}=\frac{1}{2\pi i}\int_{C}dz \left(\frac{1}{z}\right)
t_{R}\left(\frac{1}{z}\right)t(z) \nonumber \\
h=\frac{1}{2\pi iP_{out}}\int_{C}dz t_{R}\left(\frac{1}{z}\right)
\frac{dt(z)}{dz} .
\end{eqnarray}
Thus, we see that both the probability of first hitting the edge between
$j$ and $j+1$ in $n$ steps and the hitting time to reach this edge are
related to the transmission amplitude of the graph. 
 
\section{Application to example}
Let us now apply this formula to a walk on the graph discussed in  
Section 3.  The transmission amplitude for this graph is
\begin{equation}
t(\theta )=\frac{4e^{-i\theta}(1+e^{-i\phi})(1-e^{-i(4\theta +\phi)})}
{e^{-4i\theta}(1+e^{-i\phi})^{2}-(3e^{-i(4\theta +\phi)}-1)^{2}} .
\end{equation}
and
\begin{equation}
t(z )=\frac{4z^{3}(1+e^{-i\phi})(z^{4}-e^{-i\phi})}
{z^{4}(1+e^{-i\phi})^{2}-(3e^{-i\phi}-z^{4})^{2}} .
\end{equation} 
From this expression we see explicitly that $t(z)$ is analytic inside and 
on the unit circle.

Considering the case $\phi =0$ we find the simple expression
\begin{equation}
t(z)=\frac{8z^{3}}{9-z^{4}} .
\end{equation}  
Inserting this result into the expressions in the previous section, it
is straightforward to perform the integrations, and we find that
\begin{eqnarray}
q(n)& =& \left\{ \begin{array}{cr}\left(\frac{8}{9^{(n+1)/4}}\right)^{2} & 
{\rm if}\ n=3\ {\rm mod}4 \\ 0 & {\rm otherwise} \end{array} \right. 
\nonumber \\
P_{out}& = & \frac{4}{5} .
\end{eqnarray}
For small $n$, it is relatively straightforward to verify the expression 
for $q(n)$ by adding up the amplitudes for the possible paths from 
$|-1,0\rangle$ to $|2,3\rangle$ of length $n$.

Another issue that deserves mention is the existence of bound states.  In
standard quantum scattering theory, if a potential has bound states, 
these are orthgonal to all scattering states.  What we mean by a bound
state of a graph, is an eigenstate of $U$ whose support is confined to $G$.
If our graph has bound
states, then these states could not be accessed by a particle coming
in from one of the tails, e.g.\ from the left.  This follows from the
fact that the initial state is orthogonal to any bound state (a state
localized on  one of the tails is automatically orthogonal to one that
is localized in $G$), and if $\Psi_{init}$ is the initial state, and
$\Psi_{bnd}$ is the localized eigenstate with eigenvalue $\exp (-i
\theta_{b})$, we have
\begin{equation}
\langle\Psi_{bnd}|U^{n}\Psi_{init}\rangle = e^{-in\theta_{b}}\langle
\Psi_{bnd}|\Psi_{init}\rangle = 0.
\end{equation}  
This limits the walks that particles with initial states of this type can
perform, because, as the above equation shows, the state of the walk at
any time must be orthogonal to all of the bound states.

In order to find the bound states in our example we consider a 
state $|\Psi\rangle$
that is located on the four edges connecting the vertices $0$, $1A$,
$1B$, and $2$, i.e.
\begin{eqnarray}
|\Psi\rangle & = & a_{0A}|0,1A\rangle + a_{1A}|1A,2\rangle + 
b_{0A}|1A,0\rangle + b_{1A}|2,1A\rangle \nonumber \\
 & & +a_{0B}|0,1B\rangle + a_{1B}|1B,2\rangle + b_{0B}|1B,0\rangle
+b_{1B}|2,1B\rangle .
\end{eqnarray}
We then solve the equation $U|\Psi\rangle = e^{-i\theta}|\Psi\rangle$.
This equation only has a solution when $\phi =0$, and in that case the
eigenvalues are $i^{m}$, where $m=0,\ldots 3$.  The eigenstates in this
case are given by
\begin{eqnarray}
a_{0A}=\frac{1}{2\sqrt{2}} & b_{0A}=-\frac{i^{m}}{2\sqrt{2}} \nonumber \\
a_{1A}=\frac{1}{2i^{m}\sqrt{2}} & b_{1A}=\frac{-(-1)^{m}}{2\sqrt{2}}
\nonumber \\
a_{0B}=\frac{-1}{2\sqrt{2}} & b_{0B}=\frac{i^{m}}{2\sqrt{2}} \nonumber \\
a_{1B}=\frac{-1}{2i^{m}\sqrt{2}} & b_{1B}=\frac{(-1)^{m}}{2\sqrt{2}}
\end{eqnarray}
A simple basis for the space spanned by these eigenstates is
\begin{eqnarray}
|u_{1}\rangle & = & \frac{1}{\sqrt{2}}(|0,1A\rangle -|0,1B\rangle ) 
\nonumber \\
|u_{2}\rangle & = & \frac{1}{\sqrt{2}}(|1A,2\rangle -|1B,2\rangle ) 
\nonumber \\
|u_{3}\rangle & = & \frac{1}{\sqrt{2}}(|1B,0\rangle -|1A,0\rangle ) 
\nonumber \\
|u_{4}\rangle & = & \frac{1}{\sqrt{2}}(|2,1B\rangle -|2,1A\rangle ) . 
\end{eqnarray}
A state resulting from a walk starting on the tails, will be orthogonal
to the subspace spanned by these four states.  A walk starting in this
subspace will never leave the subspace, and hence, will never escape
into the tails. 

For all of the eigenvalues except those given by $i^{m}$, the corresponding
space of (unnormalized) eigenvectors is two dimensional.  It is spanned by
two scattering states, one with a incoming wave from the left, and one with
an incoming wave from the right.  These states are orthogonal.  At the 
eigenvalues $i^{m}$ the space of eigenvectors becomes three dimensional.  In
addition to the two scattering states, there is one bound state.  All of these
state are mutually orthogonal.

\section{Time reversal}
In quantum one-dimensional scattering problems, the time reversal invariance
of the Hamiltonian is used to prove that the transmission amplitude for a
wave coming from the left is the same as that for a wave coming from the
right.  Here we would like to see if something similar is possible for
quantum walks.

Our first task is to define a time-reversal operator.  The standard 
time-reversal operator is anti-unitary, does not affect positions, and
reverses velocities.  This suggests that we define the anti-unitary
operator, $\hat{T}$, where the hat is used to indicate that the operator
is anti-linear, by
\begin{equation}
\hat{T}|A,B\rangle =|B,A\rangle .
\end{equation}
Note that $\hat{T}$ is its own inverse.
For a walk to be time-reversal invariant, we want the operator that
advances the walk one step, $U$, to be taken into its inverse when
conjugated by $\hat{T}$, 
\begin{equation}
\label{invrnt}
\hat{T}U\hat{T} = U^{-1}=U^{\dagger} ,
\end{equation}
that is, conjugation with $\hat{T}$ changes the operator that moves the
walk forward one step into the one that moves it backwards one step.

Let us see what this condition implies about $U$.  Let $k$ enumerate the
vertices that are connected to $A$ by an edge (one of which is $B$).  
This set will be denoted by $\Gamma (A)$.  We find that
\begin{eqnarray}
\hat{T}U\hat{T}|A,B\rangle & = & \sum_{k}\langle A,k|U|B,A\rangle^{\ast}
|k,A\rangle
\nonumber \\
U^{-1}|A,B\rangle & = & \sum_{k}\langle A,B|U|k,A\rangle^{\ast}|k,A\rangle
\end{eqnarray}
These will be the same if
\begin{equation}
\langle A,k|U|B,A\rangle = \langle A,B|U|k,A\rangle ,
\end{equation}
so that $U$ will satisfy the condition of time-reversal invariance if for
all vertices $A$, and $ k, k^{\prime}\in\Gamma (A)$, we have
\begin{equation}
\langle A,k|U|k^{\prime},A\rangle = \langle A,k^{\prime}|U|k,A\rangle .
\end{equation}
We find that equal-transmission vertices satisfy this condition, and they
still satisfy it if phase shifters are put into the edges.  We shall
henceforth confine our attention to graphs that do satisfy this condition.

If $|\Psi\rangle$ is an eigenstate of $U$ with eigenvalue $\exp 
(-i\theta )$, and $U$ is time-reversal invariant, then
\begin{eqnarray}
\hat{T}U\hat{T}|\Psi\rangle & = & e^{i\theta}|\Psi\rangle \nonumber \\
U\hat{T}|\Psi\rangle & = & e^{-i\theta}\hat{T}|\Psi\rangle  ,
\end{eqnarray}
where we have made use of Eq.\ (\ref{invrnt}).  This implies that if
$|\Psi\rangle$ is an eigenstate of $U$, then so is $\hat{T}|\Psi\rangle$,
with the same eigenvalue.  Note that if $|\Psi\rangle$ is a bound state,
this equation implies either that $|\Psi\rangle$ is invariant under the
action of $\hat{T}$, or that the eigenvalue $\exp (-i\theta )$ is degenerate.

We can now show that the transmission coefficients from both directions are
equal.  Consider two eigenstates, each with eigenvalue $\exp (-i\theta )$,
one incident from the left, and one incident from the right
\begin{eqnarray}
\label{leftright}
|\Psi_{l} (\theta )\rangle & = & \sum_{k=-\infty}^{-1}(e^{i(k+1)\theta}
|k,k+1\rangle +r_{l}(\theta )e^{-i(k+1)\theta}|k+1,k\rangle )+|\Psi_{Gl}
(\theta )\rangle \nonumber \\ 
 & & +\sum_{k=j}^{\infty}t_{l}(\theta )e^{i(k-j)\theta }|k,k+1\rangle
\nonumber \\
|\Psi_{r}(\theta )\rangle & = & \sum_{k=-\infty}^{-1}t_{r}(\theta)
e^{-i(k+1)\theta}|k+1,k\rangle +|\Psi_{Gr}\rangle \nonumber \\
 & & \sum_{k=j}^{\infty}(e^{-i(k-j)\theta }|k+1,k\rangle +r_{r}(\theta )
e^{i(k-j)\theta }|k,k+1\rangle ) .
\end{eqnarray}
Now, consider the eigenstate
\begin{equation}
|\Psi\rangle = A|\Psi_{l}\rangle +B|\Psi_{r}\rangle .
\end{equation}
If we apply $\hat{T}$ to this state, we must obtain an eigenstate of $U$
with an eigenvalue of $\exp (-i\theta )$, which must, therefore, be a
linear combination of $|\Psi_{l}\rangle$ and $|\Psi_{r}\rangle$,
\begin{equation}
\hat{T}|\Psi\rangle = C|\Psi_{l}\rangle +D|\Psi_{r}\rangle .
\end{equation}
This implies that
\begin{eqnarray}
A^{\ast}=Cr_{l}+Dt_{r} & A^{\ast}t^{\ast}_{l}+B^{\ast}r^{\ast}_{r}=D
\nonumber \\
A^{\ast}r^{\ast}_{l}+B^{\ast}t^{\ast}_{r}=C & B^{\ast}=Ct_{l}+Dr_{r} .
\end{eqnarray}
Substituting from the second and third of these equations into the
first, gives us
\begin{equation}
|r_{l}|^{2}+t_{r}t^{\ast}_{l}=1 \hspace{1cm} r_{l}t^{\ast}_{r}
+t_{r}r^{\ast}_{r}=0 .
\end{equation}
The first of these equations, and the fact that $|r_{l}|^{2}+|t_{l}|^{2}=1$,
gives us that $t_{r}=t_{l}$.  Therefore, time-reversal invariance does
imply that the left and right transmission amplitudes are the same.

\section{Analyticity of the reflection and transmission amplitudes}
We now want to see how to find, schematically, the reflection and 
transmission amplitudes of a general graph, and then deduce some of their
properties.  In particular we want to show that they are analytic in a
region that contains the unit circle and its interior.

We begin by considering the graph $\tilde{G}$ and denote the set of oriented
edges of $\tilde{G}$ by $E$.  The space in which the walk takes place is
$\mathcal{H}=L^{2}(E)$, and this immediately presents us with a problem, 
because $|\Psi_{l}(\theta )\rangle$ (see Eq.\ (\ref{geneigen})), through which
the reflection and transmission amplitudes are defined, is not in
$L^{2}(E)$.  It can certainly be viewed as a complex-valued function on the
set $E$, $(A,B)\in E\rightarrow \langle A,B|\Psi_{l}(\theta )\rangle$,
and it is in this sense that we take the following calculations until we
have more information on the nature of the specific functions.     

The transmission and reflection amplitudes are found by solving the 
eigenvalue equation, 
\begin{equation}
U|\Psi\rangle = \exp (-i\theta )|\Psi\rangle .
\end{equation}
A direct calculation shows that this is equivalent to 
\begin{equation}
\label{eigen2}
U(|-1,0\rangle +|\Psi_{G}(\theta )\rangle )=e^{-i\theta}(|\Psi_{G}(\theta )
\rangle + r(\theta )|0,-1\rangle +t(\theta )|j,j+1\rangle ) .
\end{equation}
Now let $\mathcal{H}_{G}$ be the subspace of $\mathcal{H}$ spanned by the
oriented edges of $G$, $\mathcal{H}_{3}$ be the subspace of $\mathcal{H}$
spanned by the oriented edges $\{ |-1,0\rangle , |0,-1\rangle , 
|j.j+1\rangle \}$, $P_{G}$ be the orthogonal projection onto the space
$\mathcal{H}_{G}$, and $P_{3}$ be the orthogonal projection onto the
space $\mathcal{H}_{3}$.  In addition, define $\tilde{\mathcal{H}}=
\mathcal{H}_{3}\oplus\mathcal{H}_{G}$ and $\tilde{P}$ as the orthogonal
projection onto $\tilde{\mathcal{H}}$.  We can view the elements of 
$\tilde{\mathcal{H}}$ as column vectors of complex numbers
\begin{displaymath}
\mathbf{v}=
\left(\begin{array}{c} v_{1} \\ v_{2} \\ v_{3} \\ \mathbf{v}_{G} \end{array}
\right) ,
\end{displaymath}
where $v_{1}$ is the coefficient of $|-1,0\rangle$, 
$v_{2}$ is the coefficient of $|0,-1\rangle$, $v_{3}$ the coefficient
of $|j,j+1\rangle$ and $\mathbf{v}_{G}$ the $2N$-tuple of coefficients of the
edges of $G$ (we assume that $G$ has $N$ edges).  In this notation, 
Eq.\ (\ref{eigen2}) becomes   
\begin{equation}
\label{Geq}
e^{-i\theta}\left(\begin{array}{c}0 \\ r(\theta ) \\ t(\theta ) \\ 
\mathbf{v}_{G} \end{array}\right) = \left(\begin{array}{cc}\mathbf{X} & 
\mathbf{Y} \\ \mathbf{W} & \mathbf{G} \end{array} \right) 
\left(\begin{array}{c} 1 \\ 0 \\ 0 \\ \mathbf{v}_{G} \end{array}\right) .
\end{equation}
where
\begin{equation}
\left(\begin{array}{cc}\mathbf{X} & \mathbf{Y} \\ \mathbf{W} & \mathbf{G} 
\end{array} \right) = (\tilde{P}U\tilde{P})= \left(\begin{array}{cc}
P_{3}UP_{3} & P_{3}UP_{G} \\ P_{G}UP_{3} & P_{G}UP_{G} \end{array} \right) .
\end{equation}
In the above equation,
$X$ is a $3\times 3$ matrix, $Y$ is a $2N\times 3$ matrix, $W$ is a
$3\times 2N$ matrix, and $G$ is a $2N\times 2N$ matrix.  The second
and third columns of both $\mathbf{X}$ and $\mathbf{W}$ 
are identically zero.  We also note that the first rows of both $X$ and
$Y$ are identically zero.  

It is possible to break the solution of Eq.\ (\ref{Geq}) into two parts.  
Denote the single
nonzero column of $\mathbf{W}$ (the first one) by $\mathbf{w}$, i.e.
\begin{equation}
\mathbf{w}=\mathbf{W}\left(\begin{array}{c} 1 \\ 0 \\ 0 \\ \mathbf{0} 
\end{array}\right) ,
\end{equation}
where $\mathbf{0}$ is a $2N$-tuple of zeroes, so that $\mathbf{w}\in
\mathcal{H}_{G}$.  The first step in solving Eq.\ (\ref{Geq}) is solving the
equation in $\mathcal{H}_{G}$,
\begin{equation}
\label{Ginhom}
(e^{-i\theta}\mathbf{I}-\mathbf{G})\mathbf{v}_{G}=\mathbf{w} ,
\end{equation}
where $\mathbf{I}$ is the $2N\times 2N$ identity matrix.  Once we have  
found $\mathbf{v}_{G}$ from this equation, we insert the solution into 
Eq.\ (\ref{Geq}) in order to find $r(\theta )$ and $t (\theta )$.

We can extend the reflection and transmission amplitudes into the
complex plane by replacing $\exp (-i\theta )$ in the above equations by
$1/z$.  This will yield the functions $t(z)$ and $r(z)$.  In particular
we have that if $\mathbf{v}_{G}(z)$ is a solution to 
\begin{equation}
\label{Ginhomz}
(\mathbf{I}-z\mathbf{G})\mathbf{v}_{G}=z\mathbf{w} ,
\end{equation}
then 
\begin{eqnarray}
\label{tr}
t(z) & = & \sum_{(k,l)\in G}zY_{(j,j+1);(k,l)}v_{G;(k,l)}(z) \nonumber \\
r(z) & = & z[X_{(0,-1);(-1,0)}+\sum_{(k,l)\in G}Y_{(0,-1);(k,l)}
v_{G;(k,l)}(z)],
\end{eqnarray}
where the sums are over all edges in $G$.  

We are first interested
in determining whether $t(z)$ is analytic inside the unit circle.  This
will be the case if Eq.\ (\ref{Ginhom}) can be solved for all $|z|<1$, 
which will be true if the matrix $\mathbf{I}-z\mathbf{G}$ is
invertible for $|z|<1$.  This follows from the result that if an operator,
$T(z)$, is analytic in some region of the complex plane, so is its inverse,
$T^{-1}(z)$ provided it exists \cite{kato}.  Because $\mathbf{G}=P_{G}U
P_{G}$ we have that $\|\mathbf{G}\| \leq 1$, which implies that if $|z|<1$, 
then $\mathbf{I}-z\mathbf{G}$ is invertible.

We now have that $t(z)$ and $r(z)$ are analytic inside the unit circle, but
what about on the circle itself and a bit beyond?  To show that they are, 
we need to refine our arguments.

\begin{lemma}
\begin{enumerate}
\item $(e^{-i\theta}I-P_{G}UP_{G})|u\rangle =0$ 
if and only if $|u\rangle$ is an $L^{2}$ eigenstate of $U$ with support
in the edges of $G$.  We call these bound states. 
\item If $|w\rangle = P_{G}|-1,0\rangle$, then $\langle u|w\rangle = 0$,
where $|u\rangle$ is any bound state.
\end{enumerate}
\end{lemma}

\begin{proof}
\begin{enumerate}
\item
\begin{eqnarray}
0 & = & (e^{-i\theta}I-P_{G}UP_{G})|u\rangle  \nonumber \\
 &= & e^{-i\theta }P_{G}^{\perp}|u\rangle +e^{-i\theta }P_{G}|u\rangle 
-P_{G}UP_{G}|u\rangle ,
\end{eqnarray}
so $P_{G}^{\perp}|u\rangle = 0$ and the support of $|u\rangle$ lies in
the edges of $G$.  Therefore, $P_{G}U|u\rangle = \exp (-i\theta )
|u\rangle$, which implies that $P_{G}U|u\rangle =U|u\rangle$, and 
$|u\rangle$ is the desired eigenstate. 
\item This follows from
\begin{equation}
\langle u|w\rangle = \langle u|P_{G}U|-1,0\rangle = \langle U^{\dagger}
u|-1,0\rangle =0 .
\end{equation} 
\end{enumerate}
\end{proof}

If $e^{-i\theta}$ corresponds to a bound state, then the operator
$(e^{-i\theta}\mathbf{I}-\mathbf{G})$ is not invertible.  At the moment
this leaves us without a method of defining $t(z)$ and $r(z)$ for
values of $z$ corresponding to bound states.  We would like to show, however, 
that for these values of $z$, scattering solutions, that is
solutions to Eq.\ (\ref{Ginhom}), also exist.  These solutions can then be
used to define $t(z)$ and $r(z)$.

Let $|u_{1}\rangle ,\ldots |u_{K}\rangle$ be the eigenstates of $U$,
where $|u_{k}\rangle \in \mathcal{H}_{G}$ and $U|u_{k}\rangle =
\exp (-i\theta_{k})|u_{k}\rangle$.  Let $\mathcal{H}_{b}$ be the span
of $\{ |u_{1}\rangle ,\ldots |u_{K}\rangle\}$.  $\mathcal{H}_{b}$ is
an invariant subspace of both $U$ and $\mathbf{G}$.  Let $\mathcal{H}_{1}
=\mathcal{H}_{b}^{\perp}$ as a subspace of $\mathcal{H}_{G}$, and let
$P_{1}$ be the orthogonal projection of $\mathcal{H}_{G}$ onto
$\mathcal{H}_{1}$, and $\mathbf{G}_{1}=P_{1}\mathbf{G}P_{1}$.  
We wish to solve the equation
\begin{equation}
(\mathbf{I}_{\mathcal{H}_{1}}-z\mathbf{G}_{1})\mathbf{v}=z\mathbf{w} ,
\end{equation}
on $\mathcal{H}_{1}$. Because of the construction of $\mathcal{H}_{1}$, 
we are guaranteed that there exists an $\epsilon >0$ such that 
$(\mathbf{I}_{\mathcal{H}_{1}}-z\mathbf{G}_{1})$ is invertible on 
$\mathcal{H}_{1}$ for $|z|<1+\epsilon $.  This allows us to define
\begin{eqnarray}
|\Psi_{G}(z)\rangle & = & \mathbf{v}_{G}(z) = (\mathbf{I}_{\mathcal{H}_{1}}
-z\mathbf{G}_{1})^{-1}z\mathbf{w} \nonumber \\
r(z)& = & z(\langle 0,-1|U|-1,0\rangle +\langle 0,-1|U|\Psi_{G}(z)\rangle )
\nonumber \\
t(z)& = & z(\langle j,j+1|U|-1,0\rangle +\langle j,j+1|U|\Psi_{G}(z)\rangle ),
\end{eqnarray}
and
\begin{eqnarray}
\label{psi(z)}
|\Psi (z)\rangle & = & \sum_{k=-\infty}^{-1}z^{(k+1)}|k,k+1\rangle 
+\sum_{k=-\infty}^{-1}r(z)z^{-(k+1)}|k+1,k\rangle \nonumber \\
 & & +|\Psi_{G}(z)\rangle + \sum_{k=j}^{\infty}t(z)z^{k-j}|k,k+1\rangle .
\end{eqnarray}

We shall summarize our results in a theorem. 

\begin{theorem}
\begin{enumerate}
\item The complex-valued functions $r(z)$ and $t(z)$, and 
the function $|\Psi_{G}(z)\rangle$, which takes values in $\mathcal{H}_{1}$,
are analytic functions on the disc $|z|<1+\epsilon$. 
\item For each $z\in D=\{ z|0<|z|<1+\epsilon \}$, the punctured disc, 
$|\Psi (z)\rangle$ is an element of $L^{\infty}(E)$, and $|\Psi (z)\rangle$
is an analytic function of $z$ on the domain $D$ with values in 
$L^{\infty}(E)$. 
\item $|\Psi (z)\rangle$ is the solution to the generalized eigenvalue equation
$zU|\Psi (z)\rangle = |\Psi (z)\rangle$.
\end{enumerate}
\end{theorem}

Let us now make two brief observations.  First, 
note that the functions $t(\theta )$ and $r(\theta )$ are just the 
restrictions of $t(z)$ and $r(z)$ to the unit circle. Second, we see that 
the above theorem, along with the expressions for  $t(z)$ and $r(z)$, 
provides the necessary conditions for all of the integrals in 
Eq.\ (\ref{int0}) to vanish.

\section{Spectral results}
The formal eigenvectors $|\Psi(\theta )\rangle$, which we just constructed,
are elements of $L^{\infty}(E)$, and hence, for any $|v\rangle\in L^{1}(E)$,
the inner product $\langle\Psi (\theta )|v\rangle$ is well-defined, because
it is absolutely summable.  However, if $|v\rangle$ is merely in $L^{2}(E)$,
$\langle\Psi (\theta )|v\rangle$ is not necessarily summable.  Due to the
specific form of $|\Psi (\theta )\rangle$, this sum represents a Fourier
series with square-summable coefficients, and hence it is a function in 
$L^{2}(T)$, where $T$ is the unit circle with the usual measure $d\theta$.
We shall now exploit these two interpretations of our formulas.

Recall that $\mathcal{H}_{b}$ is the space of bound states, i.e.\ $L^{2}$
eigenfunctions of $U$.  $\mathcal{H}_{b}$ is a finite-dimensional,
$U$-invariant subspace of $\mathcal{H}$.  Each element of $\mathcal{H}_{b}$
has its support contained among the edges of $G$.  The function (vector)
$|\Psi (\theta)\rangle$ constructed in the last section is seen to be the
unique element of $L^{\infty}(E)$ such that, 
\begin{enumerate}
\item $U|\Psi (\theta )\rangle =e^{-i\theta}|\Psi (\theta )\rangle$,
\item $\langle\Psi (\theta )|-1,0\rangle = 1$, 
\item $\langle\Psi (\theta )|v\rangle = 0$ for any $|v\rangle\in
\mathcal{H}_{b}$.
\end{enumerate}
The analytic extension of $|\Psi (\theta )\rangle$ to $D=\{ z|0<|z|<
1+\epsilon \}$ is given by $|\Psi (z)\rangle$ (see Eq.\ (\ref{psi(z)})),
and a direct application of the Cauchy integral theorem to $|\Psi (z)\rangle$
implies that
\begin{equation}
\frac{1}{2\pi}\int_{0}^{2\pi}d\theta\, |\Psi (\theta )\rangle = |-1,0\rangle .
\end{equation}
If $|v\rangle\in L^{1}$, then $\langle\Psi (\theta )|v\rangle$ is a
continuous function on the circle, and 
\begin{equation}
\frac{1}{2\pi}\int_{0}^{2\pi}d\theta\, |\Psi (\theta )\rangle\langle\Psi
(\theta )|v\rangle \in L^{\infty}(E) .
\end{equation}
In particular, if $|v\rangle = |-1,0\rangle$, we have
\begin{equation}
|-1,0\rangle =  \frac{1}{2\pi}\int_{0}^{2\pi}d\theta\, |\Psi (\theta )\rangle 
\langle\Psi (\theta )|-1,0\rangle  .
\end{equation}
From this we have that
\begin{eqnarray}
U^{k}|-1,0\rangle & = & \frac{1}{2\pi}\int_{0}^{2\pi}d\theta\, e^{-ik\theta}
|\Psi (\theta )\rangle \langle\Psi (\theta )|-1,0\rangle \nonumber \\
 & = & \frac{1}{2\pi}\int_{0}^{2\pi}d\theta\, |\Psi (\theta )\rangle 
\langle (U^{\dagger})^{k}\Psi (\theta )|-1,0\rangle \nonumber \\
 & = & \frac{1}{2\pi}\int_{0}^{2\pi}d\theta\, |\Psi (\theta )\rangle 
\langle\Psi (\theta )|U^{k}|-1,0\rangle .
\end{eqnarray}
Hence, for each $|v\rangle$ in the linear span of the states 
$U^{k}|-1,0\rangle$, where $k$ is an integer, and for any trigometric 
polynomial $f$, we have the $L^{\infty}$ formulas
\begin{eqnarray}
|v\rangle & = & \frac{1}{2\pi}\int_{0}^{2\pi}d\theta\, |\Psi (\theta)\rangle
\langle\Psi (\theta )|v\rangle  \\ \label{C}
f(U)|v\rangle & = & \frac{1}{2\pi}\int_{0}^{2\pi}d\theta\, f(e^{-i\theta})
|\Psi (\theta)\rangle \langle\Psi (\theta )|v\rangle  \\ \label{D}
\langle v|f(U)|v\rangle & = & \frac{1}{2\pi}\int_{0}^{2\pi}d\theta\, 
f(e^{-i\theta})|\langle\Psi (\theta )|v\rangle |^{2} .
\end{eqnarray}
Eq.\ (\ref{C}) is an $L^{\infty}$ formula, however, Eq.\ (\ref{D}) is more
general.  It holds for any $|v\rangle$ that is a finite linear combination
of the $U^{k}|-1,0\rangle$, for $k$ an integer.  However, $\theta\rightarrow
\langle\Psi (\theta )|v\rangle$ is an $L^{2}$ function on the circle for
each $|v\rangle\in L^{2}(E)$.  Let $\mathcal{H}_{L}$ be the closure of
the linear span of the $U^{k}|-1,0\rangle$ in $\mathcal{H}=L^{2}(E)$.
$\mathcal{H}_{L}$ is a closed $U$-invariant subspace of $\mathcal{H}$,
and we see that Eq.\ (\ref{D}) holds for $|v\rangle\in \mathcal{H}_{L}$
and for any complex-valued continuous function on $T$.  Thus, by the 
Riesz-Markov theorem there exists a unique measure $\mu_{v}$, a Borel
measure on the circle, called the spectral measure associated to 
$|v\rangle$, such that
\begin{equation}
\langle v|f(U)|v\rangle = \int_{0}^{2\pi}d\mu_{v}(\theta )f(e^{-i\theta}) ,
\end{equation}
and hence,
\begin{equation}
d\mu_{v}=\frac{\langle v|\Psi (\theta )\rangle|^{2}}{2\pi}d\theta .
\end{equation}

If we define $V_{L}:\mathcal{H}_{L}\rightarrow L^{2}(T)$ by
\begin{equation}
V_{L}|v\rangle = \frac{1}{\sqrt{2\pi}}\langle \Psi (\theta )|v\rangle ,
\end{equation}
then Eq.\ (\ref{D}) implies that $\| v\| = \| V_{L}v\|$, so that $V_{L}$
is an isometry from $\mathcal{H}_{L}$ to $L^{2}(T)$, and 
$V_{L}(\mathcal{H}_{L})$ is a closed subspace of $L^{2}(T)$.  Finally, we
have 
\begin{equation}
(V_{L}UV_{L}^{-1})g(\theta )=e^{-i\theta}g(\theta ) ,
\end{equation}
for $g\in V_{L}(\mathcal{H}_{L})$ which implies that $V_{L}(\mathcal{H}_{L})$
is an invariant subspace of the multiplication operator $g(\theta )
\rightarrow \exp (-i\theta )g(\theta )$.  Therefore, we have that 
$V_{L}(\mathcal{H}_{L}) = L^{2}(T)$, and hence $V_{L}$ is a unitary operator
of $\mathcal{H}_{L}$ onto $L^{2}(T)$.  We can summarize our results to date
in the following theorem.

\begin{theorem}
Let $\mathcal{H}_{L}$ be the closed, linear, U-invariant subspace of 
$\mathcal{H}=L^{2}(E)$ generated by the states $U^{k}|-1,0\rangle$, 
for $k$ an integer.  Let $|\Psi (\theta )\rangle$ be the generalized
eigenfunctions constructed in the previous section, then for each
$|v\rangle\in \mathcal{H}_{L}$,
\begin{equation}
V_{L}|v\rangle = \frac{1}{\sqrt{2\pi}}\langle \Psi (\theta )|v\rangle ,
\end{equation}
defines a unitary operator from $\mathcal{H}_{L}$ onto $L^{2}(T)$, such
that for each $g(\theta )\in L^{2}(\theta )$,
\begin{equation}
(V_{L}UV_{L}^{-1})g(\theta )=e^{-i\theta}g(\theta ) .
\end{equation}  
Hence, we have constructed part of the spectral decomposition of $U$.
\end{theorem}

Let $\mathcal{H}_{R}$ be the closed, $U$-invariant subspace of $\mathcal{H}$
generated by the vectors $U^{k}|j+1,j\rangle$, for $k$ an integer.  In the
same way as was done in the previous section, we can construct generalized
eigenfunctions
\begin{eqnarray}
|\Psi_{r}(\theta )\rangle & = & \sum_{k=-\infty}^{-1}t_{r}(\theta)
e^{-i(k+1)\theta}|k+1,k\rangle +|\Psi_{Gr}\rangle \nonumber \\
 & & \sum_{k=j}^{\infty}(e^{-i(k-j)\theta }|k+1,k\rangle +r_{r}(\theta )
e^{i(k-j)\theta }|k,k+1\rangle ) ,
\end{eqnarray}
which satisfy the results of the last theorem of the previous section. If we
follow the arguments above, we arrive at the following theorem.

\begin{theorem}
Let $\mathcal{H}_{R}$ and $|\Psi_{r}(\theta )\rangle$ be as above.
Then for each $|v\rangle\in \mathcal{H}_{R}$,
\begin{equation}
V_{R}|v\rangle = \frac{1}{\sqrt{2\pi}}\langle \Psi_{r}(\theta )|v\rangle ,
\end{equation}
defines a unitary operator from $\mathcal{H}_{R}$ onto $L^{2}(T)$, such
that for each $g(\theta )\in L^{2}(\theta )$,
\begin{equation}
(V_{R}UV_{R}^{-1})g(\theta )=e^{-i\theta}g(\theta ) .
\end{equation}  
\end{theorem}

It has already been shown and is easy to see that $\mathcal{H}_{L}$,
$\mathcal{H}_{R}$, and $\mathcal{H}_{b}$ are mutually orthogonal,
$U$-invariant subspaces of $\mathcal{H}=L^{2}(E)$.  The final result of
this section is that these subspaces exhuast $\mathcal{H}$, and, therefore,
this completes the spectral decomposition of $U$ on $\mathcal{H}$.

\begin{theorem}
$\mathcal{H} = L^{2}(E)=\mathcal{H}_{L}\oplus \mathcal{H}_{R}\oplus 
\mathcal{H}_{b}$ .
\end{theorem}

\begin{proof}
There are many ways to prove this.  The following is in the spirit of our
original argument.  If $(\mathcal{H}_{L}\oplus \mathcal{H}_{R}\oplus 
\mathcal{H}_{b})^{\perp}=\mathcal{H}_{0}\neq 0$, then we can find a
generalized eigenfunction $|\Psi (\theta )\rangle$, such that
\begin{equation}
U|\Psi (\theta )\rangle = e^{-i\theta }|\Psi (\theta )\rangle ,
\end{equation}
where $|\Psi (\theta )\rangle$ has no overlap with $|-1,0\rangle$, 
$|j+1,j\rangle$, or $\mathcal{H}_{b}$.  Then a direct calculation shows
that $|\Psi (\theta )\rangle$ viewed as a function defined as on the 
edges, $E$, is identically zero.
\end{proof}

\section{Conclusion}
We have explored a number of properties of quantum walks on graphs.  The 
main result is that there is a connection between the S matrix of a graph
and the properties of a quantum walk on that graph.  In particular,
the probability that a quantum walk traverses a graph 
in $n$ steps and the conditional hitting time to make a walk traversing a graph
can be expressed in terms of the transmission amplitude of 
that graph.  This required studying the analyticity properties of the
reflection and transmission coefficients of the graph.
In addition, we have shown that graphs can have bound states 
that quantum walks starting on an edge leading into the graph cannot access.
We also defined what it means for a quantum walks on a graph to be 
time-reversal invariant, and used this property to show when transmission
amplitudes for particles coming from different directions will be the
same.

It should be possible to extend all of the arguments here 
to the case in which a
graph has more than two tails attached to it.  That means that it can have
multiple entrances and exits.  This will simply result in the S-matrix
of the graph having multiple channels.

\section*{Appendix A}
We want to solve the equation $U|\Psi\rangle = e^{-i\theta}|\Psi\rangle$.
Substituting in the explicit expression for $|\Psi\rangle$, we find,
by looking at the indicated edges
\begin{eqnarray}
|0,-1\rangle : & e^{-i\theta}b_{-1}=-\frac{1}{3}+\frac{2}{3}(b_{0A}+
b_{0B}) \nonumber \\
|0,1A\rangle : & e^{-i\theta}a_{0A}=\frac{2}{3}(1+b_{0B})-\frac{1}{3}
b_{0A} \nonumber \\
|1A,0\rangle : & e^{-i\theta}b_{0A}=b_{1A} \nonumber \\
|1A,2\rangle : & e^{-i\theta}a_{1A}=a_{0A} \nonumber \\
|2,1A\rangle : & e^{-i\theta}b_{1A}=-\frac{1}{3}a_{1A}+\frac{2}{3}a_{1B}
\nonumber \\
|2,3\rangle : & e^{-i\theta}a_{2}=\frac{2}{3}(a_{1A}+a_{1B}) \nonumber \\
|0,1B\rangle : & e^{-i\theta}a_{0B}=\frac{2}{3}(1+b_{0A})-\frac{1}{3}b_{0B}
\nonumber \\
|1B,0\rangle : & e^{-i\theta}b_{0B}=b_{1B}e^{i\phi} \nonumber \\
|1B,2\rangle : & e^{-i\theta}a_{1B}=a_{0B}e^{i\phi} \nonumber \\
|2,1B\rangle : & e^{-i\theta}b_{1B}=\frac{2}{3}a_{1A}-\frac{1}{3}a_{1B} .
\end{eqnarray}
Our object is to find $a_{2}$ and $b_{-1}$ from which we can find the
transmission and refelection coefficients, respectively, of the graph.
Keeping this in mind, these ten equations can readily be reduced to six
\begin{eqnarray}
e^{-i\theta}b_{-1} & = & -\frac{1}{3}+\frac{2}{3}(b_{0A}+ b_{0B}) 
\nonumber \\
e^{-i\theta}a_{0A} & = & \frac{2}{3}(1+b_{0B})-\frac{1}{3}b_{0A} 
\nonumber \\ 
e^{-3i\theta}b_{0A} & = & -\frac{1}{3}a_{0A}+\frac{2}{3}e^{i\phi}a_{0B}
\nonumber \\
e^{-2i\theta}a_{2} & = & \frac{2}{3}(a_{0A}+e^{i\phi}a_{0B})
\nonumber \\
e^{-i\theta}a_{0B}=\frac{2}{3}(1+b_{0A})-\frac{1}{3}b_{0B}
\nonumber \\
e^{-3i\theta -i\phi}b_{0B} & = & \frac{2}{3}a_{0A}-\frac{1}{3}e^{i\phi}
a_{0B} .
\end{eqnarray}
These equations can then be solved, giving us
\begin{eqnarray}
a_{2} & = & \frac{4e^{-i\theta}(1+e^{i\phi})(1-e^{-i(4\theta +\phi)})}
{e^{-4i\theta}(1+e^{-i\phi})^{2}-(3e^{-i(4\theta +\phi)}-1)^{2}}
\nonumber \\
b_{-1} & = & \frac{e^{-3i\theta}(1+e^{-i\phi})^{2}+e^{i\theta}
(3e^{-i(4\theta +\phi)}-1)(e^{-i(4\theta +\phi)}-3)} {e^{-4i\theta}
(1+e^{-i\phi})^{2}-(3e^{-i(4\theta +\phi)}-1)^{2}} .
\end{eqnarray}
Note that $a_{2}$ is just the transmission amplitude, $t(\theta )$,
for the graph,  and $b_{-1}$ is its reflection amplitude.

\section*{Appendix B}
We want to show that Eq.\ (\ref{reftrans}) is true. A graph is specified by
its set of vertices, $V$, and its set of edges, $E$. A state $|\psi\rangle
\in L^{2}(E)$ can be expressed as
\begin{equation}
|\psi\rangle = \sum_{A\in V}\sum_{(B,A)\in \omega_{A}}c_{(B,A)}|B,A\rangle .
\end{equation}
Consider a set of vertices in a graph, $S$, and its complement
$\overline{S}$.
Defining
\begin{eqnarray}
P_{S}(\psi ) & = & \sum_{A\in S}\sum_{(B,A)\in \omega_{A}}|c_{(B,A)}|^{2}
\nonumber \\
P_{\overline{S}}(\psi ) & = & \sum_{A\in\overline{S}}\sum_{(B,A)\in \omega_{A}}
|c_{(B,A)}|^{2} ,
\end{eqnarray} 
we have that 
\begin{equation}
\|\psi\|^{2}=P_{S}(\psi )+P_{\overline{S}}(\psi ) .
\end{equation}
Note that if $\|\psi\|^{2}=1$, then $P_{S}(\psi )$ is just the probability
that a particle in the state$|\psi\rangle$ will be found on one of the edges
in the set $\{ (B,A)|A\in S,\ (B,A)\in \omega_{S}\}$, with similar
considerations for $P_{\overline{S}}$, and this justifies the notation.

We can also define
\begin{eqnarray}
\tilde{P}_{S}(\psi ) & = & \sum_{A\in S}\sum_{(A,B)\in \tau_{A}}
|c_{(A,B)}|^{2} \nonumber \\
\tilde{P}_{\overline{S}}(\psi ) & = & \sum_{A\in\overline{S}}
\sum_{(A,B)\in \tau_{A}}|c_{(A,B)}|^{2} ,
\end{eqnarray}
and we then have that
\begin{equation}
\|\psi\|^{2}=\tilde{P}_{S}(\psi )+\tilde{P}_{\overline{S}}(\psi ) .
\end{equation} 
Because the operator, $U$, is unitary, we clearly have that
\begin{equation}
\label{conserve}
P_{S}(\psi )+P_{\overline{S}}(\psi )=\tilde{P}_{S}(U\psi )
+\tilde{P}_{\overline{S}}(U\psi ).
\end{equation}

We can now apply this to the state $|\Psi_{G}\rangle +|-1,0\rangle$ and
choose the set $S$ to be the vertices of $G$.  Application of $U$ to
this state gives us
\begin{eqnarray}
U(|\Psi_{G}\rangle +|-1,0\rangle )& = & e^{-i\theta }(|\Psi_{G}\rangle 
+r(\theta )|0,-1\rangle \nonumber \\
 & & +t(\theta )|j,j+1\rangle ) ,
\end{eqnarray}
and a short calculation shows that $P_{\overline{S}}(\psi )= 
\tilde{P}_{\overline{S}}(U\psi ) = 0$.  We also find that
\begin{eqnarray}
P_{S}(\psi ) & = & \|\Psi_{G}(\theta )\|^{2}+1 \nonumber \\
\tilde{P}_{S}(U\psi ) & = &  \|\Psi_{G}(\theta )\|^{2} +|r(\theta )|^{2}
+|t(\theta )|^{2} .
\end{eqnarray}
Inserting these results into Eq.\ (\ref{conserve}) Eq.\ (\ref{reftrans}) 
follows.  

We can actually do a bit more than this.  Consider Eq.\ (\ref{leftright})
in which the eigenstate for a particle coming from the left and the 
eigenstate for a particle coming from the right are given.  The result in
the previous paragraph is just
\begin{equation}
|r_{l}(\theta )|^{2}+|t_{l}(\theta )|^{2} = 1 ,
\end{equation}
and application of Eq.\ (\ref{conserve}) to the state $|\Psi_{Gr}\rangle
+|j+1,j\rangle$ gives us
\begin{equation}
|r_{r}(\theta )|^{2}+|t_{r}(\theta )|^{2} = 1 .
\end{equation}
In addition, if we have the state
\begin{equation}
|\phi\rangle = \sum_{A\in V}\sum_{(B,A)\in \omega_{A}}d_{(B,A)}|B,A\rangle ,
\end{equation}
we can define
\begin{eqnarray}
I_{S}(\phi ,\psi )& =& \sum_{A\in S}\sum_{(B,A)\in \omega_{A}}d_{(B,A)}^{\ast}
c_{(B,A)}  \nonumber \\
\tilde{I}_{S}(\phi ,\psi ) & = & \sum_{A\in S}\sum_{(A,B)\in \tau_{A}}
d_{(A,B)}^{\ast}c_{(A,B)} .
\end{eqnarray}
The fact that $U$ is unitary implies that
\begin{equation}
I_{S}(\phi ,\psi ) +I_{\overline{S}}(\phi ,\psi )=\tilde{I}_{S}(U\phi ,
U\psi )+ \tilde{I}_{\overline{S}}(U\phi ,U\psi ) .
\end{equation}
Applying this equation to the states $\phi =|\Psi_{Gl}\rangle +|-1,0\rangle$ 
and $\psi =|\Psi_{Gr}\rangle+|j+1,j\rangle$ gives us that
\begin{equation}
r_{l}(\theta )^{\ast}t_{r}(\theta )+t_{l}(\theta )^{\ast}r_{r}(\theta )
=0.
\end{equation}
What all of this has proved, is that the S-matrix for the graph, 
\begin{equation}
S(\theta )=\left(\begin{array}{cc} r_{l}(\theta ) & t_{r}(\theta ) \\
t_{l}(\theta ) & r_{r}(\theta ) \end{array}\right) ,
\end{equation}
which relates the incoming amplitudes to the outgoing amplitudes, is a
unitary matrix.

\bibliographystyle{unsrt}

\begin{thebibliography}{99}
\bibitem{shenvi} N.\ Shenvi, J.\ Kempe, and K.\ B.\ Whaley, Phys.\ 
Rev.\ A {\bf 67}, 052307 (2003) and quant-ph/0210064.
\bibitem{childs2} A.\ Childs, R.\ Cleve, E.\ Deotto, E.\ Farhi, 
S.\ Gutman, and D.\ Spielman, Proceedings of the 35th ACM Symposium
on the Theory of Computing (STOC03) (ACM Press, New York 2003), 
pages 59-68, and quant-ph/0209131.
\bibitem{davidovich}Y.\ Aharanov, L.\ Davidovich, and N. Zagury,
Phys. Rev. A {\bf 48},1687 (1993).
\bibitem{watrous}J.\ Watrous, Proceedings of the 33rd Symposium on the 
Theory of Computing (STOC01) (ACM Press, New York, 2001), p.\ 60.
\bibitem{childs} A.\ Childs, E.\ Farhi, and S.\ Gutmann, Quantum
Information Processing {\bf 1}, 35 (2002).
\bibitem{hillery} M.\ Hillery, J.\ Bergou and E.\ Feldman, Phys.\ Rev.\ A
{\bf 68}, 032314 (2003).
\bibitem{meyer} D.\ Meyer, J.\ Stat. Phys.\ {\bf 85}, 551 (1996).
\bibitem{nayak} A.\ Nayak and A.\ Vishwanath, quant-ph/0010117 (2000). 
\bibitem{aharonov} D.\ Aharonov, A.\ Ambainis, J.\ Kempe, and U.\ 
Vazirani, Proceedings of the 33rd Symposium on the Theory of Computing 
(STOC01) (ACM Press, New York, 2001), p.\ 50-59 and quant-ph/0012090 (2000).
\bibitem{yamasaki} T.\ Yamasaki, H. Kobayashi, and H.\ Imai in
C.\ Calude, M.\ J.\ Dinneen, and F.\ Pepper, editors, Unconventional
Models of Computation, Third International Conference, UMC 2002, Kobe,
Japan, October 15-19, 2002, Proceedings, volume 2509 of \emph{Lecture
Notes in Computer Science} (Springer, New York, 2002) p.\ 315-330 and
quant-ph/0205045 (2002).
\bibitem{bach} E.\ Bach, S.\ Coppersmith, M.\ Paz Goldschen, R.\ 
Joynt, and J.\ Watrous, quant-ph/0207003. 
\bibitem{mackay} T.\ D.\ Mackay, S.\ D.\ Bartlett, L.\ T.\ Stephenson,
and B.\ C.\ Sanders, J.\ Phys.\ A {\bf 35}, 2745 (2002).
\bibitem{moore} C.\ Moore and A.\ Russell in J.\ d.\ P.\ Rolim and S.\ 
Vadhan, editors, Proceedings RANDOM 2002 (Springer, New York, 2002)
p.\ 164-178 and quant-ph/0104137 (2001).
\bibitem{kempe} J.\ Kempe, quant-ph/0205083 (2002).
\bibitem{brun} T.\ A.\ Brun, H.\ A.\ Carteret, and A.\ Ambainis,
Phys.\ Rev.\ A {\bf 67}, 032304 (2003) and quant-ph/0208195.
\bibitem{kendon}V.\ Kendon and B.\ Tregenna, Phys.\ Rev.\ A {\bf 67},
042305 (2003) and quant-ph/0209005.
\bibitem{kemperev} J.\ Kempe, Contemporary Physics {\bf 44},
307 (2003) and quant-ph/0303081.
\bibitem{travaglione}B.\ C.\ Travaglione and G.\ J.\ Milburn, Phys.\ 
Rev.\ A {\bf 65}, 032310 (2002) quant-ph/0109076.
\bibitem{dur}W.\ D\"{u}r, R.\ Rausendorf, V.\ M.\ Kendon, and H.\ 
-J.\ Briegel, Phys.\ Rev.\ A {\bf 66}, 052319 (2003) and quant-ph/0203037.
\bibitem{sanders}B.\ C.\ Sanders, S.\ D.\ Bartlett, B.\ Treganna, 
and P.\ L.\ Knight, quant-ph 0207028.
\bibitem{zhao}Z.\ Zhao, J.\ Du, H.\ Li, T.\ Yang, and J.\ -W.\ Pan,
quant-ph/0212149.
\bibitem{jeong}H.\ Jeong, M.\ Paternostro, and M.\ S.\ Kim, 
quant-ph/0305003.
\bibitem{knight1}Peter L.\ Knight, Eugenio Roldan, and J.\ E.\ Sipe,
Phys.\ Rev.\ A {\bf 68}, 020301 (2003) and quant-ph/0304201.
\bibitem{knight2}Peter L.\ Knight, Eugenio Roldan, and J.\ E.\ Sipe,
Opt.\ Commun.\ {\bf 227}, 147 (2003) and quant-ph/0305165.
\bibitem{bouwmeester}D.\ Bouwmeester, I.\ Marzoli, G.\ P.\ Karman, 
W.\ Schleich, and J.\ P.\ Woerdman, Phys.\ Rev.\ A {\bf 61},
013410 (2000).
\bibitem{schmidt}A.\ G.\ M.\ Schmidt, B.\ K.\ Cheng, and M.\ G.\ 
E.\ da Luz, J.\ Phys.\ A {\bf 36}, L545 (2003).
\bibitem{farhi} E.\ Farhi and S.\ Gutmann, Phys.\ Rev.\ A {\bf 58},
915 (1998).
\bibitem{zeilinger} A.\ Zeilinger, H.\ J.\ Bernstein, D.\ M.\ 
Greenberger, M.\ A.\ Horne, and M.\ Zukowski, Controlling Entanglement
in Quantum Optics, in \emph{Quantum Control and Measurement}, H.\ 
Ezawa and Y.\ Murayama, eds. (North Holland, Amsterdam, 1993). 
\bibitem{kato} T.\ Kato, \emph{Perturbation Theory for Linear Operators}
(Springer-Verlag, New York, 1966), chapter 1.
\end{thebibliography}

\end{document}